\documentstyle[12pt,epsf]{article}
\catcode`\@=11
\def\eqnarray{\let\@currentlabel=\theequation\refstepcounter{equation}
    \global\@eqnswtrue
    \global\@eqcnt\z@\tabskip\@centering\let\\=\@eqncr
    $$\halign to \displaywidth\bgroup\@eqnsel\hskip\@centering
      $\displaystyle\tabskip\z@{##}$&\global\@eqcnt\@ne 
       \hfil${{}##{}}$\hfil
      &\global\@eqcnt\tw@ $\displaystyle\tabskip\z@{##}$\hfil 
       \tabskip\@centering&\llap{##}\tabskip\z@\cr}
\def\lefteqn#1{\hbox to 4\arraycolsep{$\displaystyle #1$\hss}}

\newcommand{\prepr}[1] {\begin{flushright}  {\bf #1} \end{flushright} \vskip
1.cm}
\newcommand{\titul}[1] {\begin{center}{\Large {\bf #1 } } \end{center}
\vskip 0.8cm}

\newcommand{\autor}[1] {\begin{center}  {\bf \lineskip .3cm #1 }
                         \end{center} }

\newcommand{\where}[1] {\begin{center}  {\normalsize \bf \it #1   } \end{center}}

\newcommand{\abstr}[1] {{\begin{center} \vskip .5cm {\bf \large Abstract
                        \vspace{0pt}} \end{center}}\begin{quote} \small #1
                        \end{quote}}
\newcounter{muni}

\begin{document}
\hbadness=10000
\pagenumbering{arabic}
\begin{titlepage}
\prepr{APCTP/97-01 \\ (Revised Version)}
\titul{\bf  Dilatonic Dark Matter\footnote{This essay received 
an ``honorable mention''
 from the Gravity Research Foundation, 1997 --- Ed. } \\  
         -- A New Paradigm --}
\autor{Y.\ M.\ Cho}
\where{Asia Pacific Center for Theoretical Physics, \\ \& }
\where{Department of Physics, College of Natural Sciences \\
Seoul National University, Seoul 151-742, Korea } 
\autor{Y.\-Y.\  Keum}
\where{ Asia Pacific Center for Theoretical Physics \\
 207-43 Cheongryangri-Dong Dongdaemun-Gu, Seoul 130-012, Korea }

\begin{center}
e-mail : yongmin@ymcho.snu.ac.kr, keum@apctp.kaist.ac.kr
\end{center}

\thispagestyle{empty}
\abstr{
We study the possibility that the dilaton plays 
the role of the dark matter of the universe.
We find that the condition for the dilaton 
to be the dark matter of the universe strongly restricts 
its mass to be around 0.5 keV or 270 MeV. 
For the other mass ranges, the dilaton
either undercloses or overcloses the universe. 
The 0.5 keV dilaton has the free-streaming distance of about 1.4 Mpc 
and becomes an excellent candidate of a warm dark matter, 
while the 270 MeV one has the free-streaming distance of about 7.4 pc 
and becomes a cold dark matter.
We discuss the possible ways to detect the dilaton experimentally.
}
\vskip 1cm

\thispagestyle{empty}
\end{titlepage}
\pagebreak

\baselineskip 22pt

\newpage
The standard big bang cosmology has been very successful in 
many ways. 
But at a deeper level the model also raises more challenges,
and may need a generalization.
This is because the standard model is based on the Einstein's
theory of gravitation, which itself may need a generalization.
Of course the Einstein's theory has been a
most beautiful and successful theory of gravitation.
But from the logical point of view
there are many indications that something 
is missing in the Einstein's theory. 
We mention just a few:

\noindent (1) The unification of all interactions 
inevitably requires the existence of a fundamental spin-zero field.
In fact all modern unified theories, from the Kaluza-Klein theory
to the superstring, contain such a fundamental scalar
field. What makes this scalar field unique is that, 
unlike others  like the Higgs field, 
it couples directly to the (trace of) energy-momentum tensor of the matter
field.
As such it should generate a new force which will 
modify the Einstein's gravitation  
in a fundamental way.

\noindent (2) The Newton's constant $G$, which is supposed to be one of the
fundamental constants of Nature, plays a crucial role in Einstein's
theory. But the ratio between the electromagnetic fine structure 
constant $\alpha_{e}$ and the gravitational fine structure constant
$\alpha_g$ of the hydrogen atom is too small to be considered natural,
$\alpha_g / \alpha_{e} \simeq 10^{-40}$.
This implies that $G$ may not be a fundamental
constant, but in fact a time-dependent parameter\cite{dirac}. 
If so, one must treat it as a fundamental scalar field which couples
to all matter fields.
Obviously this requires a drastic generalization
of Einstein's theory.

\noindent (3) In cosmology the inflation at the early
stage of evolution may be unavoidable.
But for a successful inflation we need a dynamical mechanism
which can (not only initiate but also) stop
 the inflation smoothly.
Unfortunately the Einstein's gravitation alone can not provide enough
attraction.
Again one may need a scalar field which could generate 
an extra attractive force.

All these arguments, although mutually independent, suggest the 
existence of a fundamental scalar field which we call the dilaton
which could affect the gravitation (and consequently the cosmology)
in a fundamental way.
In the following we  discuss how the dilaton comes about in the 
unified field theories, and how it could provide the dark matter of 
the universe.

Let us first consider the $(4+n)$--dimensional Kaluza-Klein
theory whose fundamental ingredient is
the $(4+n)$--dimensional metric $g_{AB}$ $(A,B=0,1,\cdots,3+n)$
\begin{equation}
g_{AB}=\left(
 \matrix{
 \tilde{g}_{\mu\nu}+ e_0 \kappa_0 \phi_{ab} A^a_{\mu} A^b_{\nu} &
  e_0 \kappa_0 A_{\mu}^a \phi_{ab} \cr
 e_0 \kappa_0 \phi_{ab} A^b_{\nu} &  \phi_{ab} \cr}
\right),
\end{equation}
where $e_0$ is a coupling constant, and $\kappa_0$ is a scale parameter
which sets the scale of the $n$-dimensional internal space.
When the metric has an $n$-dimensional  isometry $G$, one can reduce the 
$(4+n)$-dimensional Einstein's theory to a $4$-dimensional unified
theory\cite{cho,cho4}.
 Indeed with $e_0^2 \kappa_0^2 = 16 \pi G $,
$\tilde{g} = |{\rm det} ~\tilde{g}_{\mu \nu}| , 
\phi =|{\rm det}~\phi_{ab}|$, and
$\rho_{ab} = \phi^{-\frac{1}{n}}\phi_{ab}$ 
$(|{\rm det}~\rho_{ab}|=1)$,
the $(4+n)$-dimensional Einstein's theory is reduced to the following 
$4$-dimensional Einstein-Yang-Mills theory,
\begin{eqnarray}
\label{lag1}
{\cal L}_0 = -\frac{1}{16\pi G}\sqrt{\tilde{g}}
\sqrt{\phi}~\Bigl[&&\tilde{R} + \tilde{S}
+ 4\pi G\phi^{\frac{1}{n}}\rho_{ab}F_{\mu\nu}{}^aF_{\mu\nu}{}^b
-\frac{n-1}{4n}\frac{(\partial_{\mu}\phi)^2}{\phi^2} \nonumber \\
&&+\frac{1}{4}\rho^{ab} \rho^{cd}(D_{\mu}\rho_{ac})(D_{\mu}\rho_{bd})
+\Lambda+ \lambda(|{\rm det}~\rho_{ab}|-1) +\cdots
   \Bigr],
\end{eqnarray}
where $\tilde{R}$ and $\tilde{S}$ are the scalar curvature of
 $\tilde{g}_{\mu\nu}$ and $\phi_{ab}$, $\Lambda$ is the $(4+n)$-dimensional
 cosmological constant, $\lambda$ is a Lagrange multiplier.
But notice that the above Lagrangian has a crucial defect. 
First the metric 
$\tilde{g}_{\mu\nu}$ does not represent 
the Einstein's gravitation because $\tilde{g}$
does not describe the proper $4$-dimensional volume element. 
Furthermore the $\phi$-field appears with a negative 
kinetic energy, and thus can not be treated as a physical field.
To cure this defect one must perform the following conformal transformation,
and introduce the Einstein metric $g_{\mu\nu}$ and the dilaton field
$\sigma$ by \cite{cho4}
\begin{equation}
g_{\mu\nu} = \sqrt{\phi}~\tilde{g}_{\mu\nu},\hspace{5mm}
\sigma = \frac12 \sqrt{\frac{n+2}{n}}\ln \phi.
\end{equation}
With this the Lagrangian (\ref{lag1}) is written as 
\begin{eqnarray}
\label{lag2}
{\cal L}_0 = 
- \frac{\sqrt{g}}{16\pi G}\Bigl[&& R 
+ \frac12 (\partial_{\mu}\sigma)^2 
+ 4\pi G e^{\alpha\sigma}\rho_{ab} F_{\mu\nu}{}^aF_{\mu\nu}{}^b \nonumber \\
& & + S e^{-\alpha\sigma} + \Lambda e^{-\beta\sigma}
-\frac14 (D_{\mu}\rho^{ab})(D_{\mu}\rho_{ab}) 
+\lambda(|{\rm det}~\rho_{ab}|-1)
+ \cdots 
\Bigr],
\end{eqnarray}
where $R$ and $S$ are the scalar curvature of $g_{\mu\nu}$ and $\rho_{ab}$,
$\alpha$ and $\beta$ are the coupling constants given by 
$\alpha = \sqrt{(n+2)/n}$ and $\beta = \sqrt{n/(n+2)}$ .
This shows that in the Kaluza-Klein theory the dilaton appears 
as the volume element of the internal metric
which, as a component of the metric $g_{AB}$, must couple to all
matter fields.
In the superstring theory the dilaton appears 
as the massless scalar field that
the mass spectrum of the closed string must contain.   
After the full string loop expansion, the $4$-dimensional
effective Lagrangian of the massless modes has the following form 
in the string frame\cite{fra,damour} 
\begin{eqnarray}
\label{stlag1}
{\cal L}_S &=& 
-\frac{\sqrt{\tilde{g}}}{\alpha '}
\Bigl[ \tilde C_g(\varphi)\tilde{R}
+\tilde C_{\varphi}(\varphi)(\partial_{\mu}\varphi)^2 
-\frac{\alpha'}{4}\tilde C_1(\varphi) (F_{\mu\nu}{}^a )^2
+\cdots \Bigr],
\end{eqnarray}
where $\alpha '$ is the string slope parameter, 
$\tilde g_{\mu\nu}$ is the string frame metric, 
$\varphi$ is the string dilaton,
 and $\tilde C_i(\varphi)~(i= g,\varphi,1,2,3,\cdots)$ are
 the dilaton coupling
functions to various fields.
 At present their exact forms are not known 
beyond the fact that in the limit $\varphi$ goes to $-\infty$
they should admit the following loop expansion
\begin{eqnarray}
\label{scf}
\tilde C_i(\varphi)=e^{-2\varphi}+a_{i} + b_{i} e^{2\varphi}
+c_{i} e^{4\varphi} + \cdots .
\end{eqnarray}
Now introducing the Einstein metric $g_{\mu\nu}$ with a conformal 
transformation
and replacing the original dilaton field $\varphi$ with a new  
one $\sigma$, one may put the Lagrangian~(\ref{stlag1}) into the
 following standard form\cite{damour,cho1}
\begin{eqnarray}
\label{stlag2}
{\cal L}_S =
- \frac{\sqrt{g}}{\alpha'}\Bigl[
R+\frac{1}{2}(\partial_\mu \sigma)^2
- \frac{\alpha'}{4}C_1(\sigma) F_{\mu\nu}{}^a F_{\mu\nu}{}^a 
+\cdots \Bigr].
\end{eqnarray}
Notice that in the standard form
the dilaton coupling function to gravity  $\tilde C_g(\varphi)$ and
 the self coupling function $\tilde C_{\varphi}(\varphi)$ 
disappear completely with the 
redefinition of the fields. Only the coupling functions
to the other matter fields remain.

The Lagrangian (\ref{stlag2}) looks 
very much like  the Lagrangian (\ref{lag2}). In both
 cases the dilaton appears as a fundamental scalar field.
Of course there are some differences.
One  is the form of the dilatonic coupling functions to various matter
fields. In the Kaluza-Klein theory they have simple exponential
forms, whereas in the string theory their explicit forms are not known. 
Another is the mass of the dilaton.
In the Kaluza-Klein  theory the dilaton can easily acquire a mass,
but in the superstring theory it remains masslss to all orders of
perturbation\cite{fra,damour}.
But these differences may not be so serious as it appears. 
To understand this, notice that the Lagrangian~(\ref{lag2})
is valid only at the tree level.
So with the quantum correction
 in the Kaluza-Klein
theory, the difference in the dilaton coupling functions
 between the two
theories becomes insignificant. 
As for the mass of the dilaton, there is no fundamental principle
which can keep it massless, even enough the perturbative expansion 
leaves it massless in the string theory. 
So it could acquire a mass through
some unknown non-pertubative or topological mechanism.
From this one may conclude that {\em as far as the
dilaton is concerned the string theory
 and the Kaluza-Klein theory
give us practically the same effective Lagrangian, at
 least in the
low energy approximation}.
In both cases the dilaton comes in as
 the spin-zero partner of the Einstein's spin-two
 graviton. So in the unified
 theories one must take the dilatonic
modification of the Einstein's theory
 seriously, whether one likes it or not.

As this point it should also be mentioned 
 that the Brans-Dicke theory is
very similar to the above unified theories, 
as far as the dilaton is concerned\cite{cho1}.
In fact the Brans-Dicke theory can be viewed as
a 5-dimensional Kaluza-Klein theory, 
so that the unified theories could be regarded as
the ``generalized'' Brans-Dicke theories.
There are a few characteristic features 
common in all these theories: \\
1) It is the Einstein metric $g_{\mu\nu}$, 
not the string frame metric $\tilde{g}_{\mu\nu}$, 
which describes the massless 
spin-two graviton and thus
 the Einstein's gravitation\cite{cho1}.
 In fact the $\tilde{g}_{\mu\nu}$ is a strange mixture of 
the spin-two graviton and spin-zero dilaton which
does not even describe a mass
 eigenstate. This tells that, 
when one wants to compare the theory with the Einstein's gravitation,
 one must use the Einstein frame.\\
2) The unified theories could easily accommodate a successful inflation,
since the dilaton could play the role of the inflaton\cite{La,kolb}.
More significantly the dilaton could be
the dark matter of the universe, 
because in the unified cosmology 
the dilatonic matter could easily become
the dominant matter of the universe\cite{cho5}.\\
3) The dilaton describes a
``fifth force" which modifies the Einstein's gravitation\cite{cho2}. 
This implies that   
an apparant violation of the equivalence principle must take place
in the unified theories.
So an important issue in these theories is how to minimize
the modification of the  Einstein's theory 
and the violation of the equivalence
principle to an acceptable level\cite{damour,taylor}.

Now we discuss the dilatonic dark matter.
To see how the dilaton reaches the
thermal equilibrium notice that the dominant
interaction modes of the dilaton with other matter fields are the
creation(annihilation) process quark + gluon to quark + dilaton,
and the scattering process quark + dilaton to quark + dilaton. 
Normally the dilatonic coupling strength would be 
$\alpha\, m_q/m_p$, where $\alpha$ is the 
coupling constant, $m_p$ is the Planck mass, and $m_q$ is the mass of
the quarks. 
But notice that at high temperature
(at $T\gg m_q$), the coupling strength becomes $\alpha\, T/m_p $.
 With this one can easily estimate the dilaton creation (and
annihilation) cross section 
$\sigma \simeq g^2\alpha^2 (T/m_p)^2 \times 1/T^2,$
so that the creation rate $\Gamma$ is given by
\begin{equation}
\Gamma \simeq n_q\sigma v \simeq g^2 
\alpha^2\Bigl(\frac{T}{m_p}\Bigr)^2\times T.
\end{equation}
Similarly the scattering cross section $\sigma$ is given by
$\sigma \simeq \alpha^4 (T/m_p)^4 \times 1/T^2,$
with the following interaction rate $\Gamma$
\begin{equation}
\Gamma \simeq n_q\sigma v \simeq \alpha^4 
\Bigl(\frac{T}{m_p}\Bigr)^{4} \times T.
\end{equation}
On the other hand the Hubble expansion rate $H$ in
the early universe is given by
$ H \simeq T^2/m_p$.
From this we conclude that {\it the dilaton is thermally
produced from the beginning, and decouples with the other sources
at around the Planck scale with the decoupling temperature $T_d$
given by}
\begin{equation}
T_d \simeq \frac{m_p}{\alpha^{4/3}}.
\end{equation}
Notice that the dilaton decouples with the other sources at around the
same time as the graviton does. This is indeed what one would
have expected, since the dilaton is nothing but the scalar
 counterpart of the Einstein's graviton.

Once the dilaton acquires a mass, it becomes unstable and
 decays to the ordinary matter. A typical
decay process is the two photon process and the fermion pair production
process described by the following interaction Lagrangian
\begin{equation}
{\cal L}_{int}\simeq - \frac{\alpha}{4} \sqrt{16\pi G}\, \phi
F_{\mu\nu}F_{\mu\nu}
-\beta  \sqrt{16\pi G}m\,\phi\,\bar{\psi}\,\psi.
\end{equation}
For the two photon process we obtain the following life-time at
 the tree level\cite{cho-keum}
\begin{equation}
\tau_1 \simeq \frac{16}{\alpha^2}\Bigl(\frac{m_p}{\mu}\Bigr)^2
\frac{1}{\mu},
\end{equation}
where $\mu$ is the mass of the dilaton.
Similarly for the pair production we obtain
\begin{equation}
\tau_2 \simeq  \frac{1}{2 \beta^2}
\Bigl(1-4\frac{m^2}{\mu^2}\Bigr)^{-3/2}\,
\Bigl(\frac{m_p}{m}\Bigr)^2 \, \frac{1}{\mu} \, \hspace{3mm} 
\geq \hspace{2mm} \frac{5.38}{\beta^2} \Bigl(\frac{m_p}{\mu}\Bigr)^2 \,
\frac{1}{\mu}. 
\end{equation}
A more detailed calculation which includes all possible decay
 channels allowed in the standard electroweak model 
 gives us Fig.1 for the dilaton life-time 
with respect to its mass. 
Notice that here we have assumed 
$\alpha \simeq \beta \simeq 1$ for simplicity, 
but it should be  kept in mind that in reality 
the coupling constants could turn out to be much smaller\cite{cho-keum}.

To estimate how much the dilaton contributes to
 the matter density of the present universe
one must estimate the number density of the
dilaton at present time. From the entropy
conservation of the universe one
can easily estimate the present temperature $T_{\phi}$ of
the dilaton. 
Based on the standard electroweak theory
one finds
\begin{equation}
T_{\phi}\leq \Bigl(\frac{3.91}{106.75}\Bigr)^{1/3}\, 
T_0 \simeq 0.91^{\circ} K,
\end{equation}
where $T_0$ is the present temperature of the background radiation.
Notice that again this is the temperature of the graviton
at present time. From this one can estimate the number density $n_0$
 of the dilaton at present. Assuming that
 the dilaton is stable one has
\begin{equation}
n_0 = \frac{\zeta (3)}{\pi^2}\,T_{\phi}^3 \simeq 7.5\, /cm^3.
\end{equation}
But obviously, the massive dilaton can not
be stable, and the number density
of the dilaton $n(\mu)$ must crucially depend on
its mass. So for the dilaton to provide the
critical mass of the universe one must have
\begin{equation}
\rho (\mu) = n(\mu) \times \mu = n_0 \, 
e^{-t_0/\tau(\mu)}\times \mu \simeq 
10.5 \hspace{2mm}h^2 \hspace{2mm} keV/cm^3.
\end{equation}
where $t_0$ is the age of the universe,
 $\tau(\mu)$ is the life-time of the dilaton,
and $h$ is the Hubble constant (in the unit of 100Km/sec Mpc).
A numerical calculation with $t_0 \simeq 1.5 \times 10^{10} $ years 
 shows that 
{\it there are two mass ranges, 
$\mu \simeq 0.5$ keV 
or $\mu \simeq 270$ MeV,
which can make the dilaton a candidate of the dark matter in the universe.}
In Table I the interesting physical quantites are shown for
 different values of $h$.

Notice that with $h \simeq 0.6$ the mass becomes
0.5 keV or 270 MeV.
Also notice that the $\rho(\mu)$ starts from zero when $\mu = 0$ and
reaches the maximum value at 0.5 keV $< \mu < $ 270 MeV and
again decreases to zero when $\mu = \infty$. This means
that when $\mu < $ 0.5 keV or $\mu > $ 270 MeV
 the dilaton undercloses
the universe, but when 0.5 keV $ < \mu < $ 270 MeV 
it overcloses the universe. 
From this one may conclude that 
{\em the dilaton with 0.5 keV $ < \mu < $ 270 MeV 
is not acceptable because this is incompatible with the cosmology.} 
In view of the fact that the dilaton must exist in all
 the unified field theories, the above constraint on
the mass of the dilaton should provide us an important piece of
information in search of the dilaton.

Now we discuss the possibility
of the dilatonic dark matter in more detail : \\
a) $\mu \simeq $ 0.5 keV. 
In this case the available decay channel is the
  $ \gamma\gamma $ process.
So the life-time is given by $\tau \simeq 4.0 \times 10^{26}$ years,
which tells that it is almost stable. To determine whether this
dilaton could serve as a hot or cold dark matter, one
must estimate the free-streaming distance $\lambda$ of
the dilaton. The dilaton becomes non-relativistic around
 T$\simeq \mu/3\simeq 0.17$ keV,  
long before the matter-radiation equilibrium era.
In terms of time this corresponds to \cite{kolb}
\begin{equation}
t_{NR}\simeq 1.2 \times 10^7\times
\Bigl(\frac{keV}{\mu}\Bigr)^2\,
\Bigl(\frac{T_\phi}{T_0}\Bigr)^2\, sec \simeq 
1.88 \times 10^{6} \, sec.
\end{equation}
From this one obtains
\begin{eqnarray}
\lambda \simeq&& 0.16\,
\Bigl(\frac{keV}{\mu}\Bigr)\Bigl(\frac{T_\phi}{T_0}\Bigr
) \bigg[ \ln\Bigl(\frac{t_{EQ}}{t_{NR}}\Bigr) +2\bigg]\, Mpc\nonumber\\
\simeq&& 1.4 \, Mpc.
\end{eqnarray}
Certainly this is a very interesting number, 
which tells that the 0.5 keV
dilaton becomes an excellent candidate of a warm dark matter.\\
b)
$\mu\simeq 270$ MeV. 
In this case the available decay channels
are the $\gamma\gamma$,
$e^{+} e^{-}$, and $\mu^{+} \mu^{-}$ processes
(the $\nu\bar{\nu}$ processes are assumed to be negligible).
The decay processes $\gamma\gamma, \mu^{+}\mu^{-}$ are
dominant at this energy level, and have almostly the same decay width (See Table I).
The corresponding life-time is
given by $\tau\simeq 1.1 \times 10^9$ years,
 so that only a fraction of the thermal dilaton survives now. 
For this dilaton one has
\begin{equation}
t_{NR}\simeq 1.82 \times 10^{-5} \, sec,
\end{equation}
and the corresponding free-streaming distance becomes
\begin{equation}
\lambda \simeq 7.35 \, pc.
\end{equation}
Clearly  this dilaton becomes a good
candidate for a cold dark matter.

Now the important question is how one could detect the dilaton. It
seems very difficult to detect it through the dilatonic fifth force,
because the range of the fifth force would be about
$10^{-8}$ cm (for $\mu = 0.5$ keV) or 
about $10^{-13}$ cm (for $\mu = 270$ MeV).
Perhaps a more promising way is to use the two photon decay process,
which produces two mono-energetic X-rays of
 $ E \simeq 0.25$ keV or 
$E \simeq 135$ MeV with the same polarization. 
With the local halo density of our galaxy 
$\rho_{HALO}\simeq 0.3$ GeV/$cm^3$
 one can easily find the local dilaton number density to be
$\bar n\simeq 5.83 \times 10^5 /cm^3$ for $\mu = 0.5$ keV 
and $\bar n\simeq 0.11 /cm^3$ for  $\mu = 270$ MeV. 
In both cases the local velocity of the dilaton is about $10^{-3}$ c.
So it is very important to look for the above X-ray signals
from the sky
(with the Doppler broadening of $\Delta E\simeq 10^{-3}E$)
or to perform a Sikivie-type X-ray detection experiment
with a strong electromagnetic field to enhance the dilaton conversion,
although the long life-time (for $\mu = 0.5$ keV) or 
the low local number density (for $\mu = 270$ MeV)
of the dilaton could make
such experiments very difficult.
For the  $\mu$ = 270 MeV dilaton  
one could also look for the $\mu^{+}\mu^{-}$ decay process.

One might try to detect the dilaton from the accelerator experiments.
The dilaton has a clear decay signal, but the production rate should 
be very small due to the extreme weak coupling.
So one need a huge luminosity to produce the dilaton from the 
accelerators.
There are, of course, other (indirect) ways to test the existence 
of the dilaton. For example, it may be worth to look for the impacts
of the dilaton in the stellar evolution and the supernovae explosion.
We will discuss these in a separate paper\cite{cho-keum}.

\vspace*{1.0cm}

\noindent
{\em \large  Acknowledgements} \\
\indent

 The work is supported partly by the Korean Science and Engineering 
Foundation through the Center for Theoretical Physics (SNU) and by
the Ministry of Education through the Basic  Science Research 
Program (BSRI 97-2418).


\newpage 
\begin{table}[t]
\caption{
The dilatonic dark matter and its mass, decay widths, and total 
life-time for  different values of $h$.}
\begin{center}
\begin{tabular}{|c|c|c|c|c|c|}  \hline \hline
   $h$ & 0.4 & 0.5 & 0.6 & 0.7 & 0.8 
\\   \hline \hline
 $\mu$ (keV)    & 0.224 & 0.350  & 0.504 & 0.686 & 0.896  \cr
 $\tau_{tot}$ ($10^{26}$ years) 
                  & 44.2& 11.6 & 3.88 & 1.54 & 0.69  \\
\hline \hline
 $\mu$ (MeV)    & 274.8 &  272.7 & 271
.0 & 269.6 & 268.3  \cr
 $\Gamma_{\gamma\gamma}$ ($10^{-40}$ MeV) 
                  &  87.15 & 85.15 & 83.57 & 82.26 & 81.11 \cr
 $\Gamma_{e^{+}e^{-}}$ ($10^{-55}$ MeV) 
                  &  9.60 & 9.53 & 9.47 & 9.42 & 9.38  \cr
 $\Gamma_{\mu^{+}\mu^{-}}$ ($10^{-40}$ MeV) 
                  & 107.71 & 103.36 & 99.74 & 96.78 & 94.17   \cr
 $\Gamma_{tot}$ ($10^{-40}$ MeV) 
                  &  195.86 & 188.55 & 183.30 & 179.04 & 175.28  \cr
 $\tau_{tot}$ ($10^{10}$ years) 
                  & 0.107 & 0.111 & 0.114 & 0.116 & 0.119  \\
\hline \hline
\end{tabular}
\end{center}
\end{table}

\vspace{40mm}
\begin{center}
{\bf \large Figure Caption}
\end{center}
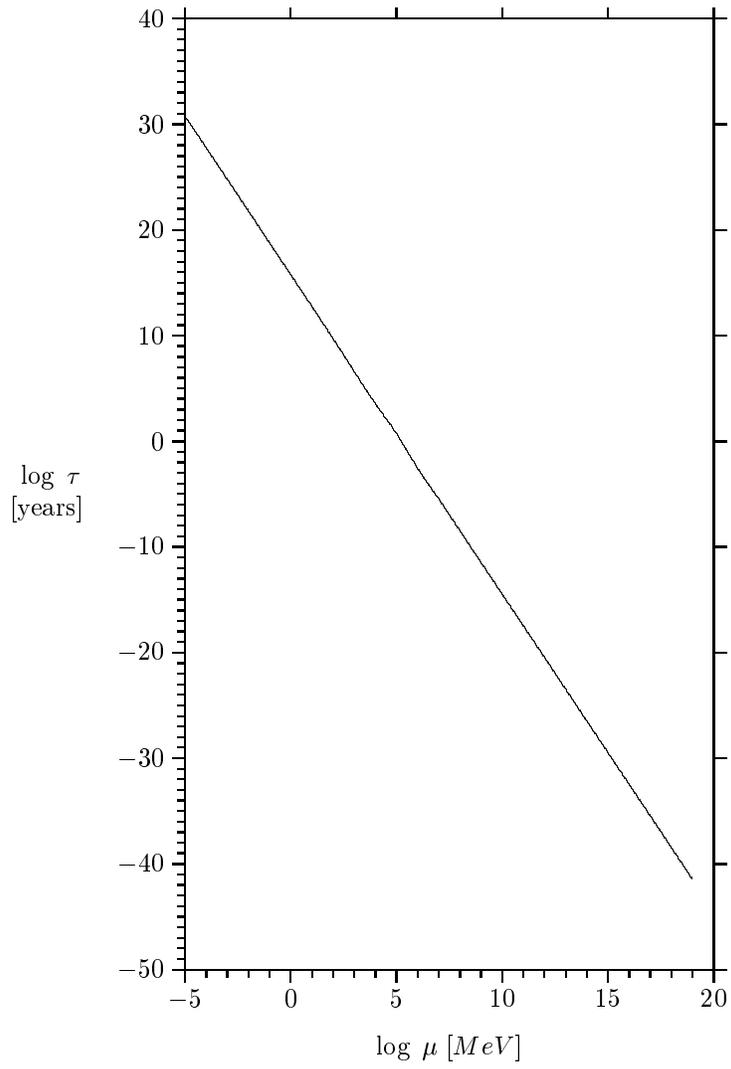
Figure 1 : The dilaton life-time versus its mass based on 
the standard electroweak model.
The result is obtained with the assumption that 
the dilatonic coupling constants to the ordinary matters 
are of the order one.

\end{document}